# Solid-state lithium-ion supercapacitor for voltage control of skyrmions

Maria Ameziane, Joonatan Huhtasalo, Lukáš Flajšman, Rhodri Mansell and Sebastiaan van Dijken

NanoSpin, Department of Applied Physics, Aalto University School of Science, P.O. Box 15100, FI-00076 Aalto, Finland

## Abstract

Ionic control of magnetism gives rise to high magneto-electric coupling efficiencies at low voltages, which is essential for low-power magnetism-based non-conventional computing technologies. However, for on-chip applications, magneto-ionic devices typically suffer from slow kinetics, poor cyclability, impractical liquid architectures or strong ambient effects. As a route to overcoming these problems, we demonstrate an LiPON-based solid-state ionic supercapacitor with a magnetic Pt/Co$_{40}$Fe$_{40}$B$_{20}$/Pt thin-film electrode which enables voltage control of a magnetic skyrmion state. Skyrmion nucleation and annihilation are caused by Li ion accumulation and depletion at the magnetic interface under an applied voltage. The skyrmion density can be controlled through dc applied fields or through voltage pulses. The skyrmions are nucleated by single 60-μs voltage pulses and devices are cycled 750,000 times without loss of electrical performance. Our results demonstrate a simple and robust approach to ionic control of magnetism in spin-based devices.

Controlling magnetism through applied voltages would allow for the creation of a new class of low-energy non-conventional computing devices. For technological applications the voltage-induced changes need to be fast, reversible and have a strong impact on the magnetic system. The ability to induce large magnetic effects at small voltages has led to an increasing interest in magneto-ionic approaches [1-3]. Previous works have shown that magnetism can be altered ionically through redox reactions [4-8], ion intercalation [9-14], or the formation of an electronic double layer at solid-ion liquid interfaces [15,16]. Devices exploiting magneto-ionics have been shown to be able to control various magnetic properties including the saturation magnetization [4-12], magnetic anisotropy [4-7,15], and Dzyaloshinskii-Moriya interaction (DMI) [17,18]. The main technological bottleneck for ionically controlled magnetism is the need to apply voltages for extended periods to create sizable effects at room temperature.

Here we take a different approach to ionic control of magnetism by creating a solid-state supercapacitor [19-21]. The large capacitance of supercapacitors is generated by ion adsorption on the electrodes leading to the creation of an electrical double layer, surface redox reactions or ion intercalation. Using a Li-enriched LiPON layer as the ion conduction layer we demonstrate fast, reversible, and durable voltage control of magnetism. In particular, we control magnetic skyrmions — topologically distinct quasiparticles of interest in magnetic data storage and non-conventional computing devices [22-27]. Previously, voltage control of skyrmions has been shown through interfacial charge modulation [17,28-33], strain transfer from piezoelectrics [34,35], and locally-applied electric fields [36]. By integrating a skyrmion-hosting magnetic thin-film structure with a supercapacitor we demonstrate nucleation and annihilation of skyrmions through sub-100 µs voltage pulses, a continuously controllable skyrmion density and the ability to extensively cycle the magnetic state without degradation. The significant improvement in the ability to control skyrmions through applied voltages demonstrated here is an important step towards technological applications, particularly neuromorphic computing [24-27].

As shown in Figure 1a, the magnetron-sputtered structure consists of an ionically conducting, 100 nm thick Li-enriched lithium phosphorous oxynitride (LiPON) layer sandwiched between a 1 nm SiN/4 nm Pt top gate electrode and a 2 nm Ta/4 nm Pt/0.9 nm CoFeB (40:40:20)/0.2 nm Pt bottom electrode. This structure is patterned into 500 µm x 500 µm crossbar junctions shown in Figure 1b (see Methods in SI). Magnetic hysteresis loops of one junction recorded under an applied bias voltage using polar magneto-optical Kerr effect (MOKE) microscopy are shown in Figure 1c, with corresponding MOKE images depicted in Figure 1d. At negative voltage the positively charged Li ions move away from the Pt/CoFeB/Pt electrode leading to a square hysteresis loop and a fully saturated film magnetization at 0 mT and +0.7 mT. The zero-voltage state shows a slanted hysteresis loop with magnetic stripe domains at 0 mT and a sparse skyrmion state at +0.7 mT. The application of a positive voltage slants the hysteresis loop further and it increases the density of the stripe domains (0 mT) and skyrmions +0.7 mT). At positive voltage the Li ions move towards the magnetic layer.

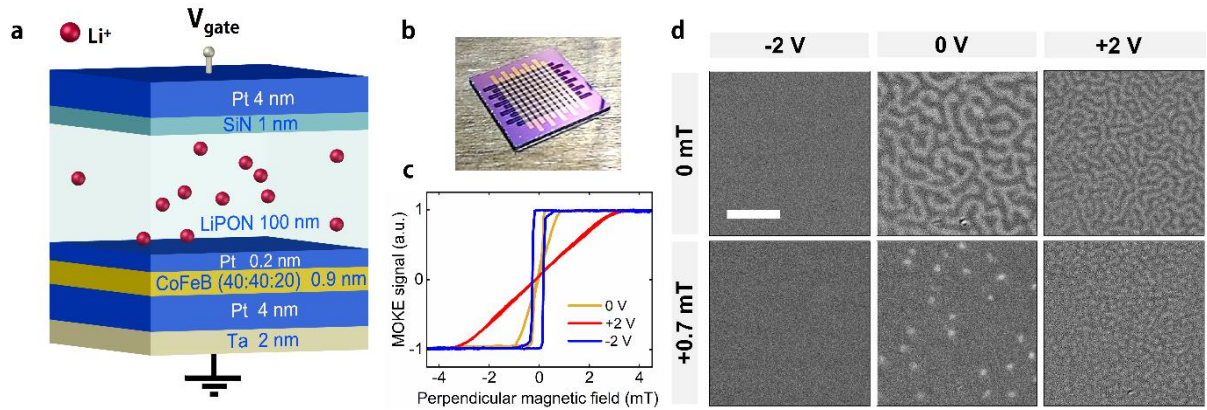

**Figure 1.** Materials system and voltage control of magnetism. (a) Schematic of the magneto-ionic heterostructure. Voltage is applied to the top gate electrode with the bottom electrode grounded. (b) Image of the crossbar sample. Both top and bottom electrodes are 500 μm wide. (c) Polar MOKE hysteresis loops recorded under 0 V, +2 V and –2 V bias voltage. (d) MOKE microscopy images for the same bias voltages under 0 mT and +0.7 mT perpendicular field. The scalebar indicates 10 μm.

To investigate control of the skyrmion density, the voltage was stepped from –1.0 V to +2.0 V and back to –1.0 V at 0.1 V intervals (Figure 2). MOKE microscopy images of the CoFeB film at selected gate voltages recorded in +0.7 mT perpendicular field are shown in Figure 2a. Starting from a saturated magnetization state at –1.0 V, inverse stripe domains form at +0.7 V, followed by the nucleation of sparse skyrmions at +0.8 V. The density of the mixed stripe and skyrmion state increases with voltage before morphing into a dense skyrmion lattice at +1.6 V. Hereafter, the skyrmion density increases further up to +2.0 V. Sweeping the voltage in the opposite direction reduces the skyrmion density gradually until all skyrmions are annihilated at –0.6 V. Figure 2b summarizes the skyrmion density during the voltage sweep. The hysteresis demonstrates the existence of a memory effect in the device, enabling access to a continuous range of skyrmion states, which is a requirement for neuromorphic devices. Besides control over skyrmion nucleation and annihilation, the gate voltage also tunes the skyrmion size (Figure 2c). The first skyrmions appearing at +0.8 V are large (~800 nm) but their size decreases continuously up to +2.0 V (~600 nm) (see Methods in SI). Sweeping the voltage in the negative direction only has a small effect on the skyrmion size. Full reversibility between a reproducible skyrmion state and no skyrmions upon repeated voltage cycling is demonstrated in Figure 2d.

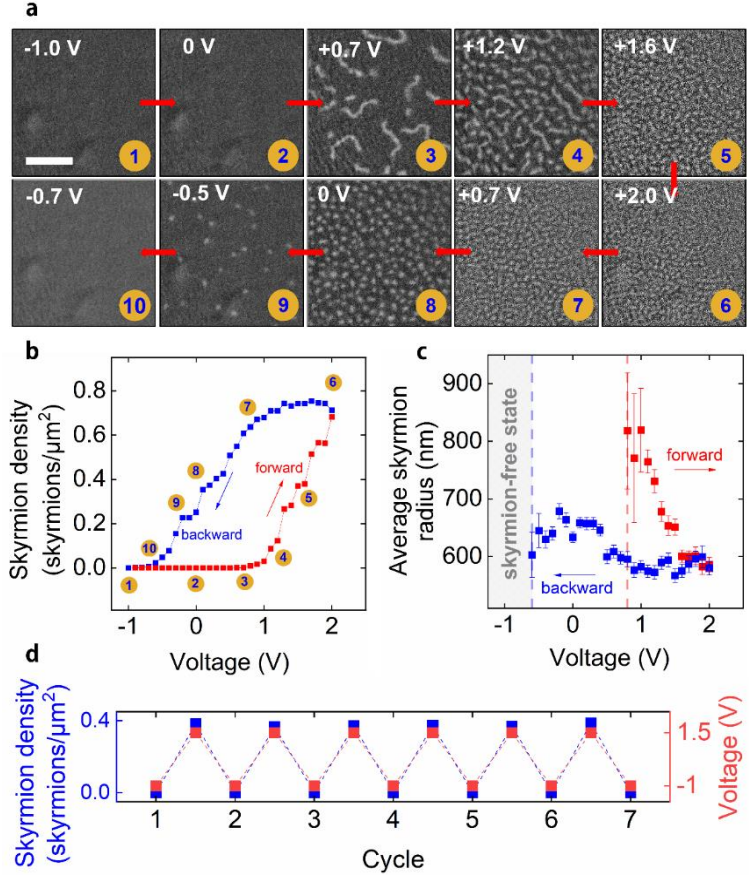

**Figure 2.** Voltage dependence of the skyrmion state. (a) Polar MOKE microscopy images recorded while sweeping the applied voltage from −1.0 V to +2.0 V and back. The perpendicular magnetic field is +0.7 mT. The scalebar corresponds to 10 µm. (b) Skyrmion density during the voltage sweep. (c) Average skyrmion radius during the voltage sweep. The error bars show the standard error of the mean size. (d) Reversible toggling of the skyrmion density by switching the voltage between −1.0 V and +1.5 V. The voltage is applied for 1 min before data collection.

For applications, devices are likely to be controlled by voltage pulses, where the response to both the application and removal of a voltage is relevant to the device operation. To investigate the decay of the skyrmion state over time at zero-bias voltage, we applied +2.0 V for 1 min to a crossbar junction followed by setting the voltage to zero. The skyrmion density as a function of time is shown in Figure 3a and MOKE microscopy images at different times are depicted in Figure 3b. The decay constant is found to be approximately 8 min.

To assess the dynamic response of our magneto-ionic device under voltage pulsing, we applied 250 ms long voltage pulses with magnitudes ranging from +1.7 V to +2.0 V at 500 ms intervals and monitored the skyrmion density over time (Figure. 3c). The device was reset to a skyrmion-free state between each series of pulses by applying −2 V for 5 s. MOKE microscopy images of the CoFeB film taken after 3000 pulses are shown in Figure 3d for four different pulse voltages. Two clear features stand out, firstly that the rate of approach to an equilibrium value is much faster at higher applied voltage and secondly that the equilibrium skyrmion density is much higher at higher applied voltage. The device shown here was cycled over 50,000 times whilst retaining the voltage control of the skyrmion state.

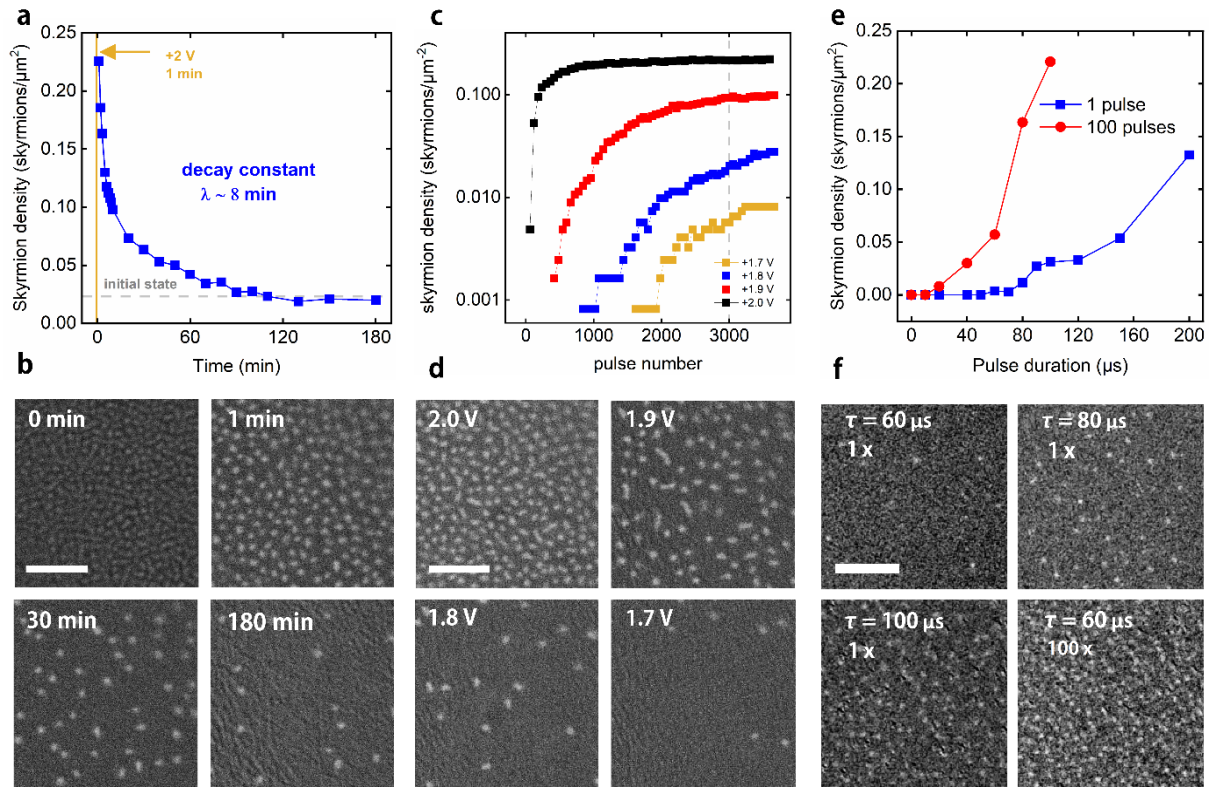

**Figure 3.** Control of skyrmions by voltage pulse number, pulse amplitude and pulse duration. (a) Time evolution of the skyrmion density at 0 V after skyrmion nucleation at +2.0 V for 1 min. (b) MOKE microscopy images recorded during the retention experiment shown in (a). The scalebar indicates 10 μm. (c) Skyrmion density as a function of voltage pulse number using a pulse duration of 250 ms with a 50% duty cycle. The pulse amplitude is varied from +1.7 V to +2.0 V. (d) MOKE microscopy images recorded after 3000 pulses for each of the applied voltages. The scalebar indicates 10 μm. (e) Skyrmion density after applying a single voltage pulse and a sequence of 100 voltage pulses to the uniform magnetization state. The amplitude of the pulse is fixed at +10.0 V and the duration of the pulse is varied. The 100-pulse sequence has a duty cycle of 10%. (f) MOKE microscopy images recorded after 60 μs, 80 μs, 100 μs for both single- and 100 pulses. The scalebar corresponds to 20 μm. All experiments used a +0.7 mT perpendicular magnetic field.

We further exploit the dependence of the skyrmion density on voltage to probe the skyrmion nucleation kinetics at shorter timescales. By applying a single pulse of +10 V, we show that the pulse width required for skyrmion nucleation can be as low as 60 μs (Figure 3e, blue curve). In these experiments, the device was reset by applying a –0.8 V gate voltage for 5 s before each voltage pulse and the skyrmion density was recorded for a few seconds after the pulse. For a sequence of 100 identical pulses the number of nucleated skyrmions increase and a pulse duration of just 20 µs is already sufficient to nucleate skyrmions (Figure 3e, red curve). MOKE microscopy images of the crossbar junction after a pulse or pulse sequence are presented in Figure 3f for pulse durations between 60 µs and 100 µs. This is to our knowledge the fastest achieved ionically induced response in a voltage-controlled magneto-ionic system at room temperature.

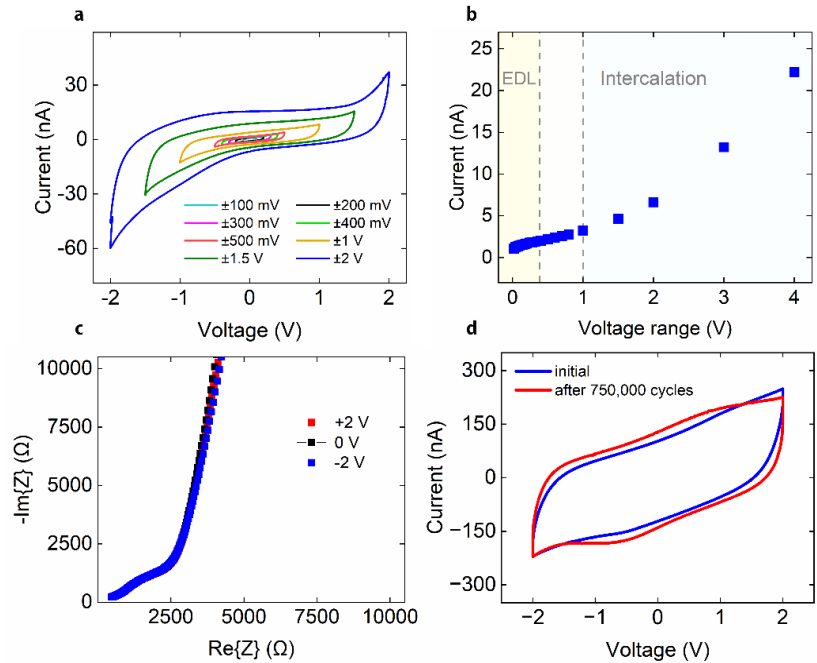

Figure 4. Electrical characterization of supercapacitor junctions. (a) Cyclic voltammograms recorded for different voltage ranges at 10 mV/s scan rate. (b) Junction current at 0 V as a function of voltage range, derived from cyclic voltammetry with a 10 mV/s scan rate. The voltage range is symmetric around 0 V. (c) Electrical impedance spectroscopy measurements on a junction using a 100 mV ac driving voltage with bias voltages of −2 V, 0 V and +2 V. (d) Cyclic voltammograms at 50 mV/s scan rate. An initial cyclic voltammogram (blue) was followed by cycling the junction between −2 V and +2 V with a period of 250 ms for 750,000 cycles, after which a second cyclic voltammogram (red) was recorded.

To understand the functioning of the devices we turn to electrical characterization. Cyclic voltammograms (CVs) of the supercapacitor structure show a largely rectangular shape with no peaks indicative of redox processes (Figure 4a). As shown in Figure 4b, for low voltage ranges the current at 0 V is a slowly increasing function of the voltage range with the junction current increasing notably for larger voltage ranges. This suggests that both electric double layer and electrochemical mechanisms are present, with the electrochemical mechanism dominating at higher voltages [8]. Given the material system it is expected that the electrochemical mechanism is intercalation of the Li ions. The capacitance of the junction is calculated to be 0.18 µF at 1 V/s, which is equivalent to a capacity of 72 µF/cm$^2$, showing large storage capability typical of supercapacitors. In Figure 4c electrical impedance spectroscopy is shown, giving a steep line at lower frequencies as expected from a capacitance-dominated device. The supercapacitor system is highly cyclable, with Figure 4d showing the CV (at 50 mV/s, giving squarer loops than in Figure 4a) before and after cycling 750,000 times with 250 ms pulses at ±2V. Moreover, our supercapacitor is intrinsically fast with a characteristic charge/discharge time of 560 μs. Figure S1 in the SI provides additional information on the electrical properties of the supercapacitor structure, including leakage current, open circuit voltage and its electrical impedance as a function of frequency.

The combination of magnetic and electrical data shows that the accumulation or depletion of Li ions at the CoFeB/Pt interface causes large changes to the magnetic state at low voltages. Values for the perpendicular magnetic anisotropy ($K_u$) and the Dzyaloshinskii-Moriya interaction constant ($D$), along with saturation magnetization ($M_s$) and exchange constant ($A_{ex}$) were estimated from a thin film sample with a similar structure (Figure S2 in the SI). $K_u$ and $D$ were found to be $9.96 \times 10^5$ J/m$^3$ and 0.74 mJ/m$^2$, respectively, which is consistent with the creation of bubble-like magnetic skyrmions in this sample at around the sizes seen in Figure 1 and Figure 2 (see Figure S3 in the SI). From our previous work [14], the insertion of Li ions at the CoFeB/Pt interface is expected to reduce the perpendicular magnetic anisotropy without reducing the magnetization [14], which reduces the energy barrier to skyrmion nucleation and stabilizes skyrmions relative to the uniform state [28] (see also SI). One interesting feature of the data in Figure 2c is the reduction in skyrmion size with increasing voltage. If the system was simply undergoing a reduction in anisotropy, then the skyrmion size is expected to increase [28,37]. Instead, a decrease in size is seen, which could either indicate that the skyrmions at lower densities are preferentially found at defect sites [38] or that the DMI is also reduced by the accumulation of Li ions at the CoFeB/Pt interface [17].

The time-dependent experiments give insight into the timescales of the phenomena. The decay time of the skyrmion state in Figure 3a corresponds to an energy barrier of around 0.38 eV, similar to that expected for the thermally activated hopping motion of Li ions within LiPON [14]. To minimize the internal electric field within the ion conduction layer there is a thermally activated redistribution of Li ions within the layer, causing the skyrmions to consequently annihilate over time. This also explains the results of the pulsed experiments in Figure 3c. Here the positive voltage pulses cause the accumulation of Li ions at the interface, which decreases the skyrmion nucleation barrier, whilst during the off state the accumulated ions decay. The concentration of interfacial Li ions increases with the number of voltage pulses until the decay in the off state balances a further increase during the on state.

For the sub-ms pulses used in Fig. 3e, there is a further effect. Now the barrier for skyrmion nucleation is lowered rapidly and then increases again as the Li accumulation decays. However, the nucleation of skyrmions occurs on a timescale longer than the voltage pulses, leading to a peak in skyrmion density around a second after the pulse (Figure S4 in the SI). Therefore, the speed of the devices is also limited by the thermally activated nucleation of the skyrmions.

Fast and durable voltage control of skyrmions in Li-ion supercapacitor structures, as shown here, offers attractive pathways to the implementation of neuromorphic devices such as synapse-based neural networks [24] and reservoir computers [25-27]. Proof-of-concepts demonstrating the suitability of skyrmion dynamics for neuromorphic computing have thus far utilized magnetic fields or electric currents to control the skyrmion state. Voltage gating of a skyrmion-hosting magnetic film provides good scalability and energy efficiency in combination with deterministic accumulation/dissipation, short-term memory, and nonlinearity. For instance, reversible nucleation and annihilation of skyrmions through the application of positive and negative voltages (Figure 2) enables the emulation of synaptic weights changes during potentiation and depression, while the decay of the skyrmion state after voltage pulsing (Figure 3a,b) provides short-term memory to temporarily store information and trigger outputs based on the time-dependent history of voltage inputs. Nonlinearity of voltage-driven skyrmion dynamics, which is another key requirement for neuromorphic processing, is demonstrated in our supercapacitors by varying the amplitude (Figure 3c,d) and duration (Figure 3e,f) of the voltage pulses. Finally, we note that the complex interplay between the dynamics of Li ion migration in the solid-state LiPON electrolyte and the ensuing nonlinear dynamics of skyrmions in the thin magnetic film offers great flexibility in the design of functional responses and further device optimization.

In summary, we have shown that skyrmions in a Pt/CoFeB/Pt thin-film structure can be created and annihilated in a fully voltage-controlled all-solid-state device via reversible Li ion migration at room temperature. The hysteretic behavior of the device with respect to the voltage sweep direction, the nonlinear effects observed as a function of voltage pulse number and pulse duration, along with the decay behavior at zero-voltage constitute properties suitable for neuromorphic device architectures. The use of a supercapacitor enables skyrmion nucleation with single voltage pulses down to 60 μs, combined with extensive cycling of the junctions. Further downscaling of the device from the 100 nm thick solid-state electrolyte used here may allow access to sub-μs functionality.

## Supporting Information

Methods and additional data, including electrical characterization, measurements to extract magnetic parameters, micromagnetic simulations of the skyrmion energy and fast voltage pulsing experiments (PDF)

## Corresponding Authors

rhodri.mansell@aalto.fi

sebastiaan.van.dijken@aalto.fi

## Author Contributions

M.A., R.M. and S.v.D. conceived the research project. M.A. grew the supercapacitor heterostructures and fabricated the crossbar junctions. M.A. conducted the electrical characterization and M.A., L.F. and R.M. performed the magnetic measurements. J.H. and R.M. conducted the micromagnetic simulations. R.M. and S.v.D. supervised the work. All authors discussed the results. M.A., R.M. and S.v.D. wrote the manuscript.

## Notes

The authors declare no competing interest.

## Acknowledgments

This work was supported by the Academy of Finland (Grant No. 316857). Lithography was performed at the OtaNano-Micronova Nanofabrication Centre, supported by Aalto University. Computational resources were provided by the Aalto Science-IT project.